\documentclass{article}
\usepackage{spconf,amsmath,graphicx}
\usepackage{xcolor}
\usepackage{comment}
\usepackage{mathtools}
\usepackage{amsfonts}
\usepackage{enumitem}
\usepackage{hyperref}

\DeclareMathOperator{\tr}{trace}

\title{Deep Deterministic Independent Component Analysis for Hyperspectral Unmixing}
\name{Hongming Li$^1$, Shujian Yu\sthanks{Contact author: yusj9011@gmail.com.}$^{2}$, Jos\'{e} C. Pr\'{i}ncipe$^1$}
\address{$^1$Computational NeuroEngineering Laboratory, University of Florida, Gainesville, FL 32611, USA\\
$^2$Machine Learning Group, UiT - The Arctic University of Norway, 9037 Troms{\o}, Norway}

\begin{document}
%
\maketitle
\begin{abstract}
We develop a new neural network based independent component analysis (ICA) method by directly minimizing the dependence amongst all extracted components. 
Using the matrix-based R{\'e}nyi's $\alpha$-order entropy functional, our network can be directly optimized by stochastic gradient descent (SGD), without any variational approximation or adversarial training. 
As a solid application, we evaluate our ICA in the problem of hyperspectral unmixing (HU) and refute a statement that ``\emph{ICA does not play a role in unmixing hyperspectral data}", which was initially suggested by~\cite{nascimento2005does}. Code and additional remarks of our DDICA is available at \url{https://github.com/hongmingli1995/DDICA}.
\end{abstract}
\section{Introduction}
\label{sec:intro}
Independent component analysis (ICA) is a statistical approach for extracting hidden factors (or sources) from a mixed signal. It has been widely applied in areas including medical image processing~\cite{calhoun2009review}, biological assays~\cite{aziz2016fuzzy}, etc. Existing ICA methods can be roughly divided into two categories by their objective functions. The first type of approaches attempt to maximize the non-Gaussianity of predicted components, since signal mixtures tend to be Gaussian distributed due to the central limit theorem~\cite{choi2005blind}. By contrast, the second type of approaches directly minimize the dependence or mutual information between different components, following the motivation that mutual information value reduces to zero if and only if all components are independent. 



Hyperspectual unmixing (HU) is a common problem in remote sensing and medical hyperspectral image analysis~\cite{bioucas2012hyperspectral}. It can be thought of as an undercomplete ICA problem where the number of mixtures is larger than sources. However, ICA-based approaches are not popular for HU mainly for two criticisms: (\textbf{C1}) the sources are sometimes mixed in a nonlinear manner, whereas most existing ICA methods assume either a linear mixing matrix (e.g., \cite{hyvarinen1999fast}) or the equivalence between the number of sources and the number of observations \cite{almeida2003misep,bach2002kernel}; and (\textbf{C2}) the HU itself is not a strict ICA problem, because the sum of abundances are required to be $1$. Due to these limitations, \cite{nascimento2005does} claims that ``\emph{ICA does not play a competitive role in unmixing hyperspectral data}".



In this paper, we challenge this point of view by developing a new deep neural network (DNN) based ICA with a softmax activation before the output and trained it with advanced information theoretic criteria. We argue that, for \textbf{C1}, the nonlinear activation functions, such as ReLU, in DNNs naturally introduce nonlinearity in a hierarchical manner. On the other hand, for \textbf{C2}, it can be simply addressed by introducing a softmax layer before the network output. 


\begin{figure}[t]
	\centering
		\includegraphics[scale=.5]{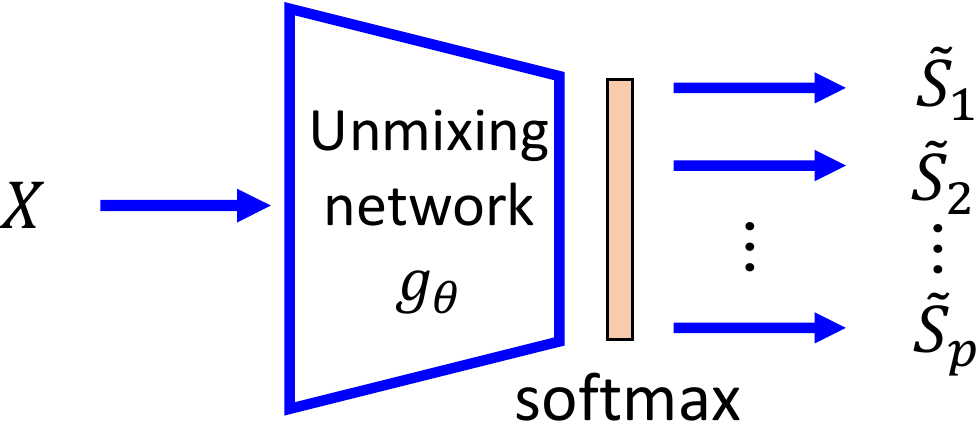}
	\caption{Architecture of our proposed deep deterministic independent component analysis (DDICA). We directly minimize the total dependence of all predicted sources $\{\tilde{S}_i\}|_{i=1}^p$ with the objective $\underset{\theta}{\arg\min}~ \texttt{TC}\left(\tilde{S}_{1},\cdots,\tilde{S}_{p}\right)$ by the matrix-based R{\'e}nyi's $\alpha$-order total correlation (TC)~\cite{Yu2019}.}
	\label{fig:DDICA}
\end{figure}

We term our method the deep deterministic ICA (DDICA), as shown in Fig.~\ref{fig:DDICA}. DDICA uses a single feed-forward neural network as the unmixing function and estimate the total mutual information non-parametrically with the recently proposed matrix-based R{\'e}nyi's $\alpha$-order total correlation~\cite{Giraldo2014,Yu2019}. DDICA can be trained directly with stochastic gradient descident (SGD) and its variants, without the necessity of variational approximation or auxiliary discriminator module. 

Our main contributions are twofold:
\begin{itemize}[leftmargin=*]
    \item Mutual information is hard to estimate. Unlike recent proposals that rely on an auxiliary discriminator network which is hard to control in practice, our DDICA can be optimized directly with SGD or its variants.
    \item Practically, we show that DDICA with a softmax layer is suitable for unmixing hyperspectral data. We show that our DDICA outperforms other traditional ICA and deep ICA methods. Our performance on HU also matches well with that achieved by state-of-the-art non-negative matrix factorization (NMF)-based approaches. This result gives new insight to the future direction of HU. 
\end{itemize}

\section{Background}
\label{sec:background}

\subsection{Traditional ICA}\label{TICA}

ICA assumes the observed random vector $\mathbf{x}=[x_1,x_2,\cdots,x_d] \\ \in R^d$ is generated by $p$ independent latent variables (also called the independent components) $\mathbf{s}=\left[s_1,s_2,\cdots,s_p\right]$ as:
\begin{equation}
    \mathbf{x} = f(\mathbf{s}),
\end{equation}
in which $f$ is a linear or nonlinear mixing function. The goal of ICA is then to recover the inverse function $f^{-1}$ (characterized by an unmixing function $g$) as well as the independent components $\left[s_1,s_2,\cdots,s_p\right]$ solely based on observations $\mathbf{x}$:
\begin{equation}
    \mathbf{\tilde{s}} = g(\mathbf{x}).
\end{equation}


The identifiability cannot be guaranteed in general nonlinear ICA problem, as there is an infinite number of solutions if the space of mixing functions is unconstrained~\cite{hyvarinen1999nonlinear}. However, some specific types of non-linear mixtures like post non-linear (PNL) ICA~\cite{taleb1999source} is solvable. In this work, we consider the case of $d>p$, which is also called the undercomplete ICA. Hyperspectral unmixing can be thought of as an undercomplete ICA problem~\cite{zhu2017hyperspectral}, as the number of mixtures $d$ is significantly larger than that of endmembers (i.e., sources) $p$.

Let the predicted sources be $\left[\tilde{s}_1,\tilde{s}_2,\cdots,\tilde{s}_p\right]$, the underlying idea of ICA can be divided into two categories.
The first type is maximizing the non-Gaussianity of predicted sources, as signal mixtures tend to be Gaussian~\cite{pati2021independent}. FastICA \cite{hyvarinen1999fast} is one of the most well-known approaches in this category. The second type is minimizing mutual information value because it reduces to zero if two random variables are independent. Notable examples in this category include Infomax~\cite{bell1995information}. Although the original Infomax assumes linearity, it can be extended to the non-linear scenario using either neural networks~\cite{almeida2003misep} or kernel tricks~\cite{bach2002kernel}. However, the number of sources is required to be the same to that of mixtures in~\cite{almeida2003misep,bach2002kernel}. Therefore they both cannot be applied to HU directly. 

\subsection{Neural ICA using parametric estimator}
DNN is a competitive functional space to represent the unmixing function $g$. 
To perform deep ICA, a differentiable mutual information (or dependence) estimator is desirable. Unfortunately, many classic mutual information estimators, e.g., $k$NN~\cite{kraskov2004estimating}, do not satisfy this requirement.
To the best of our knowledge, there are only two recent proposals on deep ICA that address this issue. Both of them use parametric approaches inspired by adversarial learning.

In~\cite{brakel2017learning}, the authors use an adversarial autoencoder to minimize JS-divergence (a substitute of MI) and reconstruction error simultaneously. 
Later, \cite{hlynsson2019learning} applies the mutual information neural estimator (MINE)~\cite{belghazi2018mutual} on ICA directly. MINE is a straightforward approach that estimates a lower bound of MI by an auxiliary neural network. 
However, these parametric approaches introduce more hyper-parameters (e.g., learning rate, number of hidden units, etc.), and their adversarial architectures, similar to GAN, are well-known for the difficulty of training and replication. 
These drawbacks motivate us to develop a new neural network based ICA with non-parametric mutual information estimators that is easy to optimize. Additional remarks about architecture differences between our ICA and~\cite{brakel2017learning,hlynsson2019learning} are further illustrated in supplementary material at \url{https://github.com/hongmingli1995/DDICA}.


\section{Methodology}\label{sec:method}


\subsection{Deep Deterministic ICA}


In this work, we use a DNN $g_\theta$, parameterized by $\theta$, as the nonlinear unmixing function. Our goal is to minimize the total dependence amongst all predicted sources $\left[\tilde{s}_1,\tilde{s}_2,\cdots,\tilde{s}_p\right]$.

One popular way to define the total dependence for $\left[\tilde{s}_1,\tilde{s}_2,\cdots,\tilde{s}_p\right]$ can be expressed as the Kullback–Leibler (KL) divergence from the joint distribution $P(\tilde{s}_1,\tilde{s}_2,\cdots,\tilde{s}_p)$ to the product of marginal distributions $P(\tilde{s}_1)P(\tilde{s}_2)\cdots P(\tilde{s}_p)$: 
\begin{equation}
\begin{aligned}
    \texttt{TC} & = D_{KL}\left[P(\tilde{s}_1,\tilde{s}_2,\cdots,\tilde{s}_p);P(\tilde{s}_1)P(\tilde{s}_2)\cdots P(\tilde{s}_p)\right],\\
    & = \left[\sum_{i=1}^p H(\tilde{s}_i)\right] - H(\tilde{s}_1,\tilde{s}_2,\cdots,\tilde{s}_p),
\end{aligned}
\label{eq:total_correlation}
\end{equation}
in which $H(\tilde{s}_i)$ is the entropy of the $i$-th source, $H(\tilde{s}_1,\tilde{s}_2,\cdots,\tilde{s}_p)$ is the joint entropy for $\left[\tilde{s}_1,\tilde{s}_2,\cdots,\tilde{s}_p\right]$. Eq.~(\ref{eq:total_correlation}) is also called the total correlation (\texttt{TC})~\cite{watanabe1960information}, and has been recently applied for disentangled representation learning~\cite{chen2018isolating} and the understanding to the dynamics of learning of DNNs~\cite{yu2021measuring}.

An alternative way to define the total dependence is by the following formula:
\begin{equation}
    \texttt{D}
    = \sum_{i=1}^p I(\tilde{s}_i;\tilde{\mathbf{s}}_{-i}),
\label{eq:dual_correlation}
\end{equation}
in which $I$ denotes mutual information and $\tilde{\mathbf{s}}_{-i}$ refers to the set of all sources except the $i$-th source. 

Eq.~(\ref{eq:dual_correlation}) has been used in~\cite{hlynsson2019learning} to train a ICA network, and the (lower bound) of $I(\tilde{s}_i;\tilde{\mathbf{s}}_{-i})$ can be evaluated by MINE. However, the poor assumptions of MINE (requires absolute continuity) and its parametric nature leads to training difficulty. Similar to GAN, numerous ``hyper-parameters", e.g., structure of estimator neural networks, learning rates, etc, need to be selected much carefully in the implementation.

Unlike~\cite{hlynsson2019learning} that requires the estimation of $p$ mutual information values in high-dimensional space, we directly optimize Eq.~(\ref{eq:total_correlation}) for simplicity. Moreover, instead of using MINE or adversarial training strategy which is hard to control in practice, we compute entropy and joint entropy terms in Eq.~(\ref{eq:total_correlation}) directly from data by the matrix-based R{\'e}nyi's $\alpha$-order entropy functional, without any variational approximation or distributional assumption. 

Suppose there are $N$ samples\footnote{One can understand $N$ as the mini-batch size in network training.} for the $i$-th predicted source, i.e., $\tilde{s}_{i} = [\tilde{s}_{i}^1, \tilde{s}_{i}^2, \cdots,\tilde{s}_{i}^N]$ where the subscript denotes view index and superscript denotes sample index. A Gram matrix $K_i \in R^{N \times N}$ can be obtained by computing $K_i(n,m) = \kappa(\tilde{s}_{i}^n, \tilde{s}_{i}^m)$, where $\kappa$ is a infinite divisible kernel~\cite{bhatia2006infinitely} which is usually assumed to be Gaussian. The entropy of $\tilde{s}_{i}$ can be expressed by~\cite{Giraldo2014}:
\begin{equation}\label{eq:renyi_entropy}
H_{\alpha}(\tilde{s}_{i}) = H_{\alpha }(A_i) = \frac{1}{1-\alpha }\log_{2}[\sum_{n=1}^{N}\lambda _{n}(A_i)^{\alpha}],
\end{equation}
where $A_i=K_i/\tr(K_i)$ is the normalized Gram matrix and $\lambda _{n}(A_i)$ denotes $n$-th eigenvalue of $A_i$.

Further, the joint entropy for $\{\tilde{s}_i\}_{i=1}^p$ is defined as~\cite{Yu2019}:
\begin{equation}\label{eq:renyi_joint_entropy}
H(\tilde{s}_1,\tilde{s}_2,\cdots,\tilde{s}_p) = H_{\alpha }(\frac{A_1\circ A_2\cdots \circ A_p}{\tr(A_1\circ A_2\cdots \circ A_p)}).
\end{equation}

The \texttt{TC} in Eq.~(\ref{eq:total_correlation}) can be directly obtained with Eqs.~(\ref{eq:renyi_entropy}) and (\ref{eq:renyi_joint_entropy}). 
The differentiability of this matrix-based Renyi's entropy estimator has been proved and empirically validated in~\cite{yu2021measuring}. In practice, the automatic SVD is embedded in mainstream deep learning APIs including TensorFlow and PyTorch. In this sense, our DDICA enjoys a simple and tractable objective $\min \texttt{TC}$, which can be directly trained with SGD or its variants.

\subsection{Application in Hyperspectral Unmixing (HU)}

Hyperspectual unmixing (HU) can be thought of as an undercomplete ICA problem where $d\gg p$. Mixtures $\mathbf{x}$ are recorded by a camera with $d$ different spectral channels, and our aim is to separate them into $p$ classes. The DDICA achieves this goal by simply changing the number of output units.
However, deep learning methods are usually sensitive to weight initialization~\cite{sutskever2013importance}. Unfortunately, such dependence to weight initialization is also observed in our experiments on HU with both MINE and our DDICA, i.e., the predicted sources of hyperspectral images are sometimes partially correct. To alleviate this problem, we train $g_\theta$ with different weight initialization independently by $T$ times. These neural networks generate $p \times T$ predicted sources. We finally cluster the predicted sources using $k$ means ($k = p$) and view the clustering centers as predicted sources. In practice, one can also select $k$ slightly larger than $p$.


\section{Experiment}
\label{sec:exp}



In this section, we conduct experiments on both synthetic and real-world hyperspectral data to demonstrate the advantages of our DDICA. We compare DDICA with a set of popular references.
\textbf{FastICA}~\cite{hyvarinen1999fast} is the most well-known ICA method based on maximizing non-Gaussianity by kurtosis. 
\textbf{InfomaxICA}~\cite{bell1995information} follows the idea of minimizing mutual information. It maximizes joint entropy of predicted components as an alternative. 
ICA based on mutual information neural estimator (\textbf{MINE})~\cite{belghazi2018mutual,hlynsson2019learning} estimates a lower bound of mutual information values by an auxiliary network. 
\textbf{MISEP} \cite{almeida2003misep} is a post non-linear (PNL) extension of InfomaxICA but cannot be directly applied to undercomplete ICA, i.e., HU. 
Additionally, we also compare the performance of DDICA with that of~\textbf{min-vol NMF}~\cite{leplat2019minimum}. This is a recently proposed NMF-based approach and outperforms several state-of-the-art (SOTA) NMF-based algorithms in HU~\cite{ekanayake2020constrained}.  




\subsection{Validation on Synthetic PNL Dataset}
We first test all competing ICA methods on actual $16$kHz audio signals~\cite{kabal2002tsp} but use synthetic PNL mixing function. We select two speech signals (FA01\_01 and MA01\_01) plus a uniform noise as our sources. Three $1$D noisy sources are first mixed linearly by a $3\times 3$ mixing matrix and then processed by $tanh(x)$, $(x+x^3)/2$, $e^x$, respectively. We set the batch size to be $2,000$, and results are normalized using Z-score.

\begin{figure}[htb]
\begin{minipage}[b]{0.24\linewidth}
  \centering
  \centerline{\includegraphics[width=2.4cm]{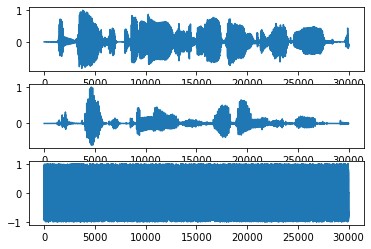}}
  \centerline{(a) Sources}\medskip
\end{minipage}
\hfill
\begin{minipage}[b]{.24\linewidth}
  \centering
  \centerline{\includegraphics[width=2.4cm]{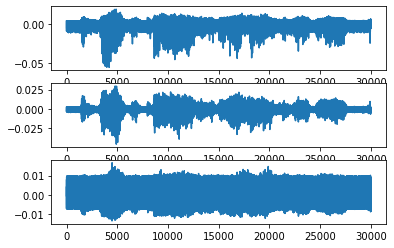}}
  \centerline{(b) FastICA}\medskip
\end{minipage}
\hfill
\begin{minipage}[b]{0.24\linewidth}
  \centering
  \centerline{\includegraphics[width=2.4cm]{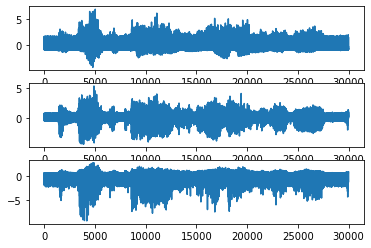}}
  \centerline{(c) Infomax ICA}\medskip
\end{minipage}

\begin{minipage}[b]{0.24\linewidth}
  \centering
  \centerline{\includegraphics[width=2.4cm]{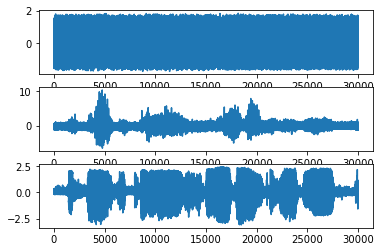}}
  \centerline{(d) MISEP}\medskip
\end{minipage}
\hfill
\begin{minipage}[b]{.24\linewidth}
  \centering
  \centerline{\includegraphics[width=2.4cm]{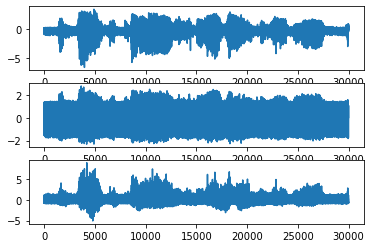}}
  \centerline{(e) MINE}\medskip
\end{minipage}
\hfill
\begin{minipage}[b]{0.24\linewidth}
  \centering
  \centerline{\includegraphics[width=2.4cm]{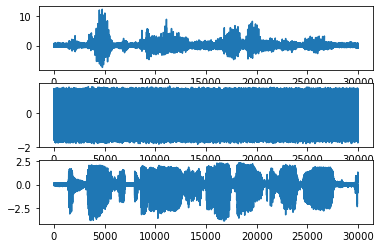}}
  \centerline{(f) Ours}\medskip
\end{minipage}
\caption{Predicted sources of different ICA methods.}
\label{pnl}
\end{figure}

The $g_\theta$ in our DDICA is a four-hidden-layer DNN with ReLU activation in the first three layers and linear activation in the last layer. It is followed by a differentiable whitening layer~\cite{schuler2019gradient} to avoid trivial solutions. Notice that, we do not add a softmax layer here, because there is no constrains on the sum of magnitude of predicted components.  The order $\alpha$ is $0.75$ and the kernel width $\sigma$ is $0.1584$ according to Silverman’s rule~\cite{Silverman2018}. Experiments are repeated for 10 times and the average scores are recorded. As demonstrated in Fig.~\ref{pnl} and Table~\ref{tab.pnl}, our DDICA outperforms other linear (FastICA, Infomax) and non-linear ICAs (MISEP, MINE). 
Besides, compared with MINE, our network is much easier to optimize (see supplementary material for details).




\begin{table}[h]
\begin{tabular}{cccccc}
Methods & FastICA & Infomax & MISEP        & MINE   & Ours            \\ \hline
$\left| \rho \right|$     & 0.7132  & 0.6646     & \underline{0.9133} & 0.7690 & \textbf{0.9527}
\end{tabular}
\caption{Separation performance in terms of absolute correlation coefficient $\left| \rho \right|$. The best performance is in bold and the second-best performance is underlined.}
\label{tab.pnl}
\end{table}


\begin{table}[htb]
\begin{tabular}{cccccc}
Method  & FastICA & Infomax & \begin{tabular}[c]{@{}c@{}}min-vol\\ NMF\end{tabular} & MINE   & Ours            \\ \hline
Soil    & 0.2942  & 0.4198  & \underline{0.1945}                                       & 0.3319 & \textbf{0.1524} \\
Tree    & 0.4632  & 0.4426  & \underline{0.2248}                                       & 0.3828 & \textbf{0.1948} \\
Water   & 0.4704  & 0.4913  & \textbf{0.1528}                                       & 0.3911 & \underline{0.2401} \\ \hline
Average & 0.4093  & 0.4512  & \textbf{0.1907}                                       & 0.3686 & \underline{0.1958}

\end{tabular}
\caption{Unmixing performance in terms of RMSE on Sansom dataset. The best performance is in bold and the second-best performance is underlined}
\label{tab.sam}
\end{table}

\begin{table}[htb]
\begin{tabular}{cccccc}
Methods & FastICA & Infomax & \begin{tabular}[c]{@{}c@{}}min-vol \\ NMF\end{tabular} & MINE   & Ours            \\ \hline
Asphalt & 0.2502  & 0.3203  & \underline{0.1699}                                        & 0.2995 & \textbf{0.1548} \\
Grass   & 0.3133  & 0.3333  & \textbf{0.1690}                                        & 0.3281 & \underline{0.1860}          \\
Tree    & 0.3712  & 0.4177  & \textbf{0.1264}                                        & 0.3404 & \underline{0.1927}          \\
Roof    & 0.4299  & 0.4458  & \underline{0.2573}                                        & 0.3978 & \textbf{0.2359} \\
Dirt    & 0.5163  & 0.4883  & 0.4075                                                 & \underline{0.3961} & \textbf{0.3225} \\ \hline
Average & 0.3762  & 0.4011  & \underline{0.2228}                                                 & 0.3544 & \textbf{0.2163}
\end{tabular}
\caption{Unmixing performance in terms of RMSE on Urban dataset. The best performance is in bold and the second-best performance is underlined.}
\label{tab.urban}
\end{table}




\begin{figure}[htb]

\begin{minipage}[b]{0.48\linewidth}
  \centering
  \centerline{\includegraphics[width=4cm]{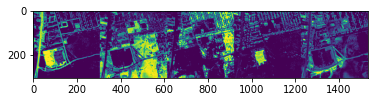}}
  \centerline{(a)  Ground Truth}\medskip
\end{minipage}
\hfill
\begin{minipage}[b]{0.48\linewidth}
  \centering
  \centerline{\includegraphics[width=4cm]{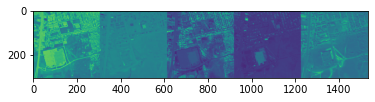}}
  \centerline{(b) FastICA}\medskip
\end{minipage}

\begin{minipage}[b]{.48\linewidth}
  \centering
  \centerline{\includegraphics[width=4.0cm]{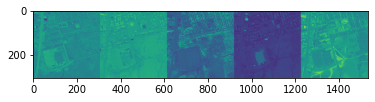}}
  \centerline{(c) Infomax ICA}\medskip
\end{minipage}
\hfill
\begin{minipage}[b]{0.48\linewidth}
  \centering
  \centerline{\includegraphics[width=4.0cm]{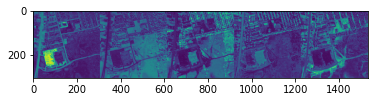}}
  \centerline{(d) MINE}\medskip
\end{minipage}

\begin{minipage}[b]{.48\linewidth}
  \centering
  \centerline{\includegraphics[width=4.0cm]{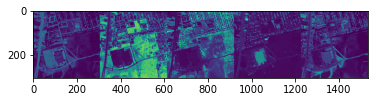}}
  \centerline{(e) min-vol NMF}\medskip
\end{minipage}
\hfill
\begin{minipage}[b]{0.48\linewidth}
  \centering
  \centerline{\includegraphics[width=4.0cm]{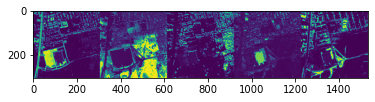}}
  \centerline{(f) Ours (10 units)}\medskip
\end{minipage}

\caption{Abundance maps extracted by ICA methods on Urban.}
\label{fig:urban}
\end{figure}

\vspace*{-5mm}
\subsection{Results of Hyperspectral Unmixing}\label{HU_result}
We select two real-world hyperspectral datasets: Samson and Urban. Each pixel of Samson is recorded by $156$ spectral channels, i.e., $d = 156$, whereas Urban contains $162$ channels. The number of sources $p$ of Samson and Urban is $3$ and $5$, respectively. In hyperspectral unmixing, the first step is usually the determination of the number of sources using techniques like virtual dimensionality algorithm~\cite{chang2004estimation}. However, the aim of this work is to show the advantages of our ICA over other existing ICA methods. Therefore, we assume the true number of sources is known.

The network $g_{\theta}$ is a six-hidden-layer DNN with ReLU activation for the first five layers and softmax activation in the last layer. The $\alpha$ of our estimator is $1.01$ to approximate Shannon entropy. The kernel width $\sigma$ is fixed to be $0.1$. We trained $1,000$ sets of parameters with different initialization to obtain $1,000 \times p$ unmixed abundances and then apply $k$-means. Clustering centers are treated as unmixing results.




Tables~\ref{tab.sam} and \ref{tab.urban} compare the root mean square error (RMSE) between true and predicted sources (see~\ref{fig:urban}). We repeat the experiments $10$ times and average the scores. From the table we observe that our method outperforms others ICA-based methods and achieves comparable results to the NMF-based one.  Compared to Samson dataset, there are more classes and details in Urban datasets. Hence our method cannot achieve perfect results if number of output channels $p$ equals to the true number of classes. 
To address this problem, we apply a strategy inspired by unsupervised neural classification algorithms \cite{ji2019invariant}. We increase the number of outputs units of the neural networks to approximate an over-classification scenario. Then $k$ means is applied to over-dimensioned unmixed maps. Fig. \ref{fig:urban} (f) illustrates results with 10 units.
We conduct the experiments on Urban dataset with five different number of output units: $[5,6,7,8,9,10]$. The corresponding RMSE are $[0.2453,0.2410,0.2274,0.2184,0.2201,0.2163]$, respectively.
There is a negative correlation between RMSE and the number of output channels, and it is also reported in \cite{nascimento2005does}. 
\section{Conclusion}
In this paper, we propose a novel deep neural network-based ICA without parametric estimation of mutual information. There are only two hyper-parameters in our mutual information estimation, i.e., order of R{\'e}nyi's entropy, kernel size. Hence our framework is easier to train compared to variational approaches. We validate our model on PNL and hyper-spectral unmixing datasets. Our method outperforms other ICA-based approaches and achieves comparable results to the SOTA NMF-based ones. 

\section{Acknowledgment}
This work was funded in part by the U.S. ONR under grants N00014-18-1-2306, N00014-21-1-2324, N00014-21-1-2295, the DARPA under grant FA9453-18-1-0039, and the Research Council of Norway (RCN) under grant 309439.

\section{Supplementary Material}
\label{sec:intro}

\subsection{Structure of Deep ICA Approaches}

To the best of our knowledge, there are only two recent proposals on deep ICA that address this issue. Both of them use parametric approaches inspired by adversarial learning.

\begin{figure}[htp]
	\centering
		\includegraphics[scale=.5]{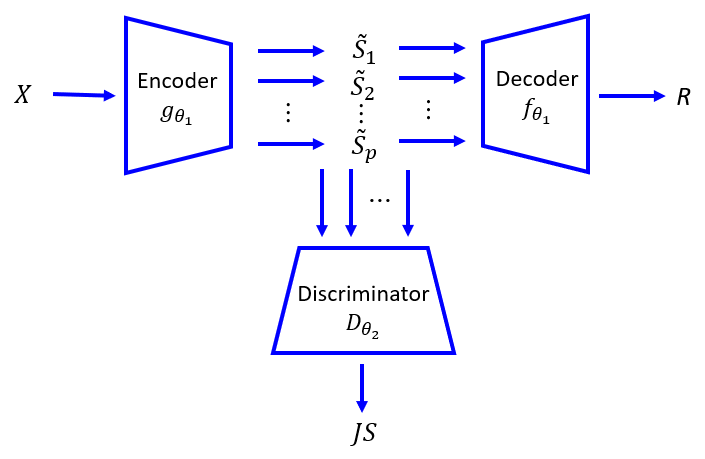}
	\caption{Architecture of the adversarial autoencoder ICA. The autoencoder minimizes the JS while the discriminator maximizes it.}
	\label{AEICA}
\end{figure}

In~\cite{brakel2017learning}, the authors use an adversarial autoencoder to minimize JS-divergence (a substitute of MI) and reconstruction error simultaneously. The detailed structure is shown in Fig. \ref{AEICA}, where $R$ denotes reconstruction error, $JS$ denotes JS-divergence. There are three deep neural modules, i.e., encoder (unmixing function), decoder and discriminator.
The autoencoder is updated by minimizing $JS+\lambda R$, where $\lambda$ is a hyper-parameter, while the discriminator is updated by maximizing $JS$.

\begin{figure}[htp]
	\centering
		\includegraphics[scale=.5]{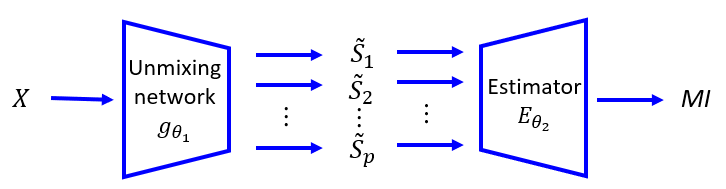}
	\caption{Architecture of the MINE-based ICA. The encoder (Unmixing function) minimizes the MI while the estimator maximizes it.}
	\label{MINE}
\end{figure}

 Mutual information neural estimator (MINE)~\cite{belghazi2018mutual} is a more straightforward approach which estimates MI directly.  As demonstrated in Fig. \ref{MINE}, where $MI$ denotes mutual information, the MINE architecture has two neural modules. The objective function of MINE is a lower bound of MI. The encoder learns to minimize this lower bound while estimator maximizes it. Our empirical study use MINE as a baseline representing neural ICA approach. We use the codes of \cite{hlynsson2019learning} which apply MINE on ICA directly.

\begin{figure}[t]
	\centering
		\includegraphics[scale=.5]{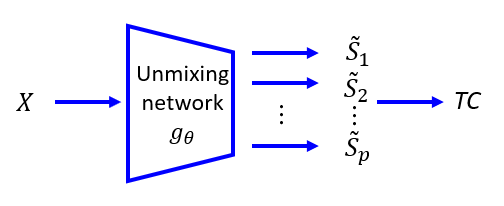}
	\caption{Architecture of our proposed deep deterministic independent component analysis (DDICA). We directly minimize the total dependence of all predicted sources $\{\tilde{S}_i\}|_{i=1}^p$ with the objective $\underset{\theta}{\arg\min}~ \texttt{TC}\left(\tilde{S}_{1},\cdots,\tilde{S}_{p}\right)$ by the matrix-based R{\'e}nyi's $\alpha$-order total correlation (TC)~\cite{Yu2019}.}
	\label{fig:DDICA}
\end{figure}

Unlike parametric independence estimation approaches, we propose to exploit a matrix-based non-parametric Renyi's entropy estimator \cite{renyi1961,Giraldo2014,Yu2019} for MI estimation. We utilize TC, a variant of MI, as our objective function and a single deep neural network as our unmixing function. As shown in Fig. \ref{AEICA} \ref{MINE} and \ref{fig:DDICA}, our architecture has only one neural module which promises much less parameters.

Besides, adversarial learning is hard to train because it introduce more hyper-parameters (learning rate, number of hidden units and layers of discriminator, etc.). In contrast, there are only two hyper-parameters in our mutual information estimation, i.e., order of R{\'e}nyi's entropy, kernel size. Hence our framework is easier to train compared to variational approaches.

\subsection{Less Hyper-parameters, Easier Training}

In this section, we conduct experiments to show the simplicity of hyper-parameters selection of our architecture and compares it with the MINE-based ICA.   

\begin{figure}[htb]

\begin{minipage}[b]{0.48\linewidth}
  \centering
  \centerline{\includegraphics[width=4cm]{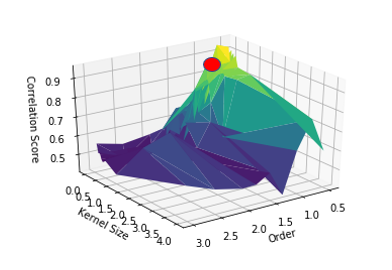}}
  \centerline{(a)  DDICA}\medskip
\end{minipage}
\hfill
\begin{minipage}[b]{0.48\linewidth}
  \centering
  \centerline{\includegraphics[width=4cm]{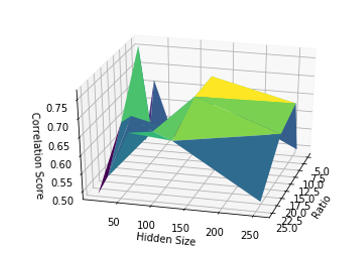}}
  \centerline{(b) MINE}\medskip
\end{minipage}

\caption{Hyper parameter space of DDICA and MINE. The red point indicates the 1.01 for order and Silverman rule for kernel size}
\label{hyper}
\end{figure}

We validate it using the PNL data shown in the paper. We train our model with different order of R{\'e}nyi's entropy $\alpha$, kernel size $\sigma$ and plot the 3D figure of it (See the left panel of Fig. \ref{hyper}). The z axis is the absolute correlation coefficient $\left| \rho \right|$. Normally, we can first set the order to be 1.01 to approximate Shannon entropy and apply Sliverman rule \cite{Silverman2018} for kernel size, that is $N^{-1/(4+p)}$, where $p$ is the dimension of the latent vector, $N$ can be considered as the batch size. We set the batch size to be 2000, and the $\sigma$ based on Sliverman rule should be $0.2187$. The point $(1.01, 0.2187)$ is demonstrated using a red point in the the left panel of Fig. \ref{hyper}, which is very close to the global optimal of hyper-parameter selection $(0.75, 1.584)$. Therefore, we claim that there is a simple hyper-parameters searching policy for our architecture. We can start from 1.01 and Sliverman rule, and then search a small space near the start point to obtain best performance. 

On the contrary, neural estimation has more hyper-parameters mainly including learning rate, hidden layers, hidden units in each layer, corporation of two neural networks. In our experiment, we fix the learning rate to be 0.0001, the number of hidden layers to be 2 but change the hidden units in each layer and updating ratio to plot the 3D figure. The updating ratio indicates how many times we update discriminator after we update the encoder for a single time. As demonstrated in the right panel of Fig. \ref{hyper}, there are multiple peaks in the hyper-parameter space, and we don't have a start point policy for our searching. It indicates that we have to search a large space for best performance.

In summary, the major advantages of DDICA compared to variational approaches are twofold:
\begin{itemize}[leftmargin=*]
    \item We have much less parameters in our architecture.
    \item We have much less hyper-parameters in our mutual information estimator, and a simple searching policy make it easy find the optimal of hyper-parameters selection.
\end{itemize}

\small
\bibliographystyle{IEEEbib}
\bibliography{strings,refs}

\end{document}